\documentstyle[twocolumn,prl,aps,epsfig]{revtex}
\renewcommand{\vec}[1]{{\mathbf{#1}}}
\newcommand{\beq}{\begin{eqnarray}} 
\newcommand{\eeq}{\end{eqnarray}} 

\begin{document} 
\draft 
 
\title 
{Local Dynamics and Strong Correlation Physics I: 
1D and 2D Half-filled Hubbard Models}
\author{Tudor D. Stanescu and Philip Phillips}

%
\address
{Loomis Laboratory of Physics\\
University of Illinois at Urbana-Champaign\\
1100 W.Green St., Urbana, IL, 61801-3080}

%

\address{\mbox{ }}
\address{\parbox{14.5cm}{\rm \mbox{ }\mbox{ }
We report on a non-perturbative approach to the 1D and 2D Hubbard models that
is capable of recovering both strong ($U\gg t$) and weak-coupling ($U\ll t$) limits,
with $U$ the on-site Coulomb repulsion and $t$ the kinetic energy.  
Dynamical corrections to the electron self-energy
in the single particle
Green function are explicitly included by expanding in terms of 
the 16
eigenstates that characterise two nearest neighbour sites.
We first show that even when $U$ is much smaller than the 
bandwith, the Mott-Hubbard gap never closes at half-filling
in both 1D and 2D.  Consequently, the Hubbard model at half-filling
is always in the strong-coupling non-perturbative regime.
For both large and small $U$, we find that the population of nearest-neighbour
singlet states approaches a value of order unity as $T\rightarrow 0$
as would be expected for antiferromagnetic
order.  We also find that the double occupancy is a smooth
monotonic function of $U$ and approaches the anticipated
non-interacting limit of $1/4$ as $U\rightarrow 0$ and vanishes
as $U\rightarrow\infty$.  Finally,
we compute the heat capacity ($C(T,U)$) for both 1D and 2D.
Our results for 1D at moderate to high temperatures are in quantitative 
agreement with those
of the exact Bethe ansatz solution, differing by no more
than 1$\%$.   In addition,
we find that in 2D, the $C(T,U)$ curves vs $T$ for different
values of $U$ exhibit a universal crossing point at two characteristic 
temperatures, $T\approx 1.7t\pm0.1t$ and $T\approx 0.4\pm 0.1t$ as is seen
universally in Hubbard models and
experimentally in a wide range of strongly-correlated systems such
as $^3He$, $UBe_3$, and $CeCu_{6-x}Al_x$. The success of this method 
in recovering well-established results that stem fundamentally
from the Coulomb interaction suggests that local dynamics are
at the heart of the physics of strongly correlated systems.
 }}
\address{\mbox{ }}
\address{\mbox{ }}

\maketitle

\section{Introduction}

In both the weak and strong coupling regimes, 
the Hubbard model at half-filling
is expected to be an antiferromagnet at $T=0$. 
The strong-coupling argument relies on
an isomorphism between the half-filled Hubbard model in the 
limit that the on-site Coulomb repulsion, $U$, exceeds the 
hopping integral, $t$, and a
Heisenberg 
antiferromagnet.  As a consequence, the corresponding ground state is 
antiferromagnetic and the energy scale for the gap between the upper 
and lower Hubbard bands is set by the energy cost for double occupancy,
$\Delta\approx U$.  In the opposite or weak-coupling limit, $U\ll t$, 
perturbation theory
predicts that a van-Hove singularity induces a spin-density
wave producing a gap that is exponentially small in the Coulomb repulsion;
that is, $\Delta\approx te^{-2\pi\sqrt{t/U}}$.  While the perturbative
argument cannot be extended into the strong-coupling regime where the 
$t/U$ mapping to a perfect Heisenberg antiferromagnet applies, continuity
between the two regimes suggests that antiferromagnetism persists for any non-zero value
of $U$.  Further, this argument would also suggest that the Hubbard gap
never closes for any non-zero value of $U$.  Consequently, the half-filled
Hubbard model is always in the strong-coupling regime.  However, no exact
results are known. In fact, while numerical simulations\cite{white}
support an antiferromagnet at half-filling, several mean-field
arguments suggest otherwise.  Dating back to the pioneering work of Mott\cite{mott} and Brinkman and 
Rice\cite{br}, numerous calculations
on the 2D\cite{weakcoupling,caste} or the $D=\infty$\cite{dinfty,dmft}
half-filled Hubbard models suggest that 
whenever $U$ is much smaller than the bandwidth, $W=8t$, the Hubbard gap 
closes and a metallic phase ensues.  Contrastly, Anderson\cite{anderson} has
argued that almost certainly the half-filled 2D Hubbard model is non-perturbative
as in the 1D case\cite{lw}, thereby possessing a discontinuity only at $U=0$. At 
the heart of the non-perturbative nature of the Hubbard model\cite{anderson} is
the projective mismatch between the low-energy physical subspace and the
``anti-bound states'' which form in 2D for any non-zero value of $U$. 
Antiferromagnetism follows necessarily as a corollary from the break-down of
perturbation theory.  

In this paper we re-examine this problem using an approach that is capable
of spanning the weak and strong-coupling regimes.  Our approach
is based on the Hubbard operators which exactly diagonalize the
interaction part of the Hubbard Hamiltonian. Consequently, the Hubbard
operators are tailor-made to access the strong-coupling regime,
$U\gg t$.   Rather than work
in the static approximation in which quantum fluctuations are ignored
leading to infinitely sharp upper and lower Hubbard bands, we include the dynamical corrections which
lead to broadening of the spectral features.  As in the work of 
Matsumoto and Mancini\cite{mm,mm1}, we focus on the dynamics associated
with two neighbouring sites.   The dominant dynamics appear to be governed by
spin-fluctuations which lead to singlet-triplet excitations. From our analysis,
we 
conclude that the Hubbard gap never closes and the
2D Hubbard model
at half-filling is always in the strong-coupling regime.  
Our results then corroborate
those of a recent improvement\cite{dca} on dynamical mean-field 
theory\cite{dmft} in which the momentum-dependence of the self-energy 
is explicitly included\cite{dca} at particular points in the
Brillouin zone.  To determine the validity of the approach we use here,
we study as well the 1D half-filled Hubbard model as 
exact results are known from Bethe ansatz\cite{bethe}.  
As expected, we find that the Hubbard gap persists even in the 
weak coupling regime.  In addition, we find that our results
for the heat capacity are in perfect agreement with those from
Bethe ansatz in the temperature range where the Coulomb interaction
dominates the physics, that is, moderate to high temperatures.
Finally, we show that we recover the well-established
universal crossing\cite{georg,dm,voll} of $C(T,U)$ vs $T$ for various values of $U$
that is seen experimentally in a wide range of strongly-correlated
systems such as $^3He$\cite{he}, $CeCu_{6-x}Al_x$\cite{ce}, 
$Nd_{2-x}Ce_xCuO_4$\cite{nd}, and $UBe_{3}$\cite{ube}. The success
of our approach suggests that local dynamics lead to many of the features
of strongly-correlated electronic systems.

\section{Dynamical Green Function Approach}

The starting point of our analysis is the on-site Hubbard model
\beq\label{HHam}
H = -\sum_{i,j,\sigma} t_{ij}c_{i\sigma}^{\dagger}c_{j\sigma} + 
U\sum_{i} n_{i\uparrow}n_{i\downarrow}
\eeq
where $t_{ij}=t$ if $(i,j)$ are nearest-neighbour sites and zero otherwise.
Rather than working with the original electron operators,
we use the Hubbard operators
$\eta_{i\sigma}=c_{i\sigma}n_{i-\sigma}$ and 
$\xi_{i\sigma}=c_{i\sigma}(1-n_{i-\sigma})$ as these operators exactly
diagonalise the interaction term. In terms of 
the Hubbard operators, $c_{i\sigma}= \eta_{i\sigma}+\xi_{i\sigma}$.  While 
the interaction term is now simplified
in this basis, the Hubbard operators do not obey standard Fermi statistics,
making impossible any diagrammatic approach based on Wick's theorem.
However, the equation of motion approach has been
 demonstrated\cite{compo,roth,tudor2} to 
offer an alternative to the diagrammatic expansion. Consider the two-component basis
\beq\label{psi}
\psi_\sigma(i)=\left(\begin{array}{l}
\xi_{i\sigma}\\\eta_{i\sigma}
\end{array}\right)
\eeq
and its associated Green function $G(i,j,t,t')=
\langle\langle \psi_{i\sigma};\psi^\dagger_{j\sigma}\rangle\rangle=
\theta(t-t')\langle \{\psi_{i\sigma}(t),\psi^\dagger_{j\sigma}(t')\}\rangle$
where $\{A,B\}$ is the anticommutator and $\langle\cdots\rangle$ is the thermal
average.  The equations of motion for the Hubbard operators,
\beq
j_i(t)=i{\partial\psi_i\over\partial t}=E_0\psi_i+\delta j^0_i
\eeq
will of course contain a contribution which is linear in the Hubbard basis
and in addition new terms, $\delta j^0_i$ which contain operators
that lie outside the Hubbard basis.   Ideally,
if such operators are included in the Hubbard basis, then
the non-linear contributions can be minimized.  However,
such a procedure is necessarlily cumbersome.  Instead of enlarging the
basis, we project $\delta j^0_i$ onto the Hubbard basis using
the Roth\cite{roth} projector,
\beq\label{roth}
{\cal P}(O)=\sum_{ln}\langle\{O,\psi_l^\dagger\}
\rangle I_{ln}^{-1}\psi_n
\eeq
which projects any operator $O$ onto the Hubbard operator basis,
where $I(\vec k)=FT\langle\{\psi_\sigma(i,t),\psi^\dagger_\sigma(j,t\}\rangle$,
and $FT$ denotes the time and space Fourier transform.  
This projector is particularly
useful because it allows us to recast the equations
of motion for the Hubbard operators 
\beq
j_i(t)=E_0\psi_i+{\cal P}(\delta j^0_i)+\delta j_i=E\psi_i+\delta j_i
\eeq
in terms of the renormalized energy matrix 
\beq
E(\vec k)=E_0+FT\langle\{\delta j^0_{i\sigma},\psi^\dagger_{l\sigma}\}\rangle I^{-1}(\vec k)
\eeq
and a correction $\delta j_i=\delta j^0_i-{\cal P}(\delta j^0_i)$.
Clearly, ${\cal P}(\delta j_i)=0$. 
The formal
solution for the Fourier transform of the Green function
\beq\label{fgf}
G(\vec k,\omega)=\frac{1}{\omega-E(\vec k)-\delta m(\vec k,\omega)}I(\vec k)
\eeq
contains the dynamical self-energy,
\beq\label{se}
\delta m(\vec k,\omega)=FT\langle R \{\delta j_{i\sigma}(t),\delta 
j^\dagger_{l\sigma}(t')\}\rangle_I
\eeq
where $R$ denotes retarded
and $I$ the irreducible part.
For a paramagnetic phase, the overlap matrix 
\beq
I=\left(\begin{array}{cc}
1-\frac{n}{2}& 0\\ 
0&\frac{n}{2}\\
\end{array}
\right)=\left(\begin{array}{cc}
I_1& 0\\ 
0&I_2\\
\end{array}
\right)
\eeq
is explicitly diagonal.  The weights which appear along the diagonal represent
the contribution from the lower and upper Hubbard bands, respectively.
Note, they sum to unity.  This feature coupled with the fact that the dynamical
corrections vanish when $U=0$ guarantees that we recover the correct
non-interacting limit.

The primary operational hurdle with any analytical approach to the Hubbard
model is the
evaluation of the dynamical self energy.  
In the static approximation\cite{compo,roth,tudor2}, 
the self-energy is dropped, and the Green function reduces to the pole structure
\beq\label{fgf2}
G(\vec k,\omega)=\frac{1}{\omega -E(\vec k)}I(\vec k)
\eeq
where $E(\vec k)$ defines the energy bands.  At this
level of theory, the Hubbard bands are sharp as the Green function
has a pole at the energy of each band.  As our focus, however, is on
the closing of the Hubbard gap, the pole approximation is inadequate and the
broadening in the bands arising from the self energy is crucial.  In the context
of the composite operator approach, Mancini and Matsumoto\cite{mm} have
 developed
a real-space scheme for computing the dynamical corrections to the 
static approximation.  To implement this procedure, we rewrite the dynamical
correction
\beq
\delta m(\vec k,\omega)&=&Dm(\vec k,\omega)\left(\begin{array}{cc}
1& -1\\ 
-1& 1\\
\end{array}
\right)\\
&\equiv &FT\langle R\delta j_i(t)\delta j^\dagger_l(t)\rangle\hat{\vec K}\nonumber
\eeq
in terms of the $2\times 2$ matrix $K$.
Because 
$Dm(\vec k,\omega)$ cannot be evaluated exactly, we seek a systematic
way of calculating the dynamical corrections.  The simplest approach
would be to consider the single-site approximation.  Such an approximation
is in the spirit to the $d=\infty$\cite{dinfty} methods in which
the self-energy is momentum independent. An improvement would be to consider
the dynamics associated with two sites as proposed by Mancini
 and Matsumoto\cite{mm,mm1}.
Evaluation of the self-energy over succesively larger clusters
would lead to an exact determination of the dynamical corrections.
Hence, we write the dynamical corrections as a series
\beq
Dm(x,x')=\delta_{x,x'}Dm_0(x,x')+\sum_a\delta_{x+a,x'}Dm_1(x,x')
+\cdots\nonumber
\eeq
in increasing cluster size.
Here, $x$ and $x'$ are neighbouring sites and $a$ indexes
all nearest-neighbour sites.  In the two-site approximation, the series is 
truncated at the level
of on-site, $Dm_0$, and nearest-neighbour, $Dm_1$ contributions.
In Fourier space, the dynamical corrections can be written
as,
\beq
Dm(\vec k, \omega)\approx Dm_0(\omega)+\alpha(\vec k)Dm_1(\omega)
\eeq
with
\beq
Dm_0(\omega)&=&\frac{1}{4}FT\langle R\{\delta j(t),\delta j^\dagger (t')\}\rangle,\nonumber\\
Dm_1(\omega)&=&\frac{1}{4}FT
\langle R\{\delta j(t),\delta j^{\prime\dagger} (t')\}\rangle,
\eeq
where the factor of $1/4$ arises from the coordination number
 on a square lattice,
$\alpha(\vec k)=(\cos k_x+\cos k_y)/2$, and $\delta j$ and 
$\delta j^{\prime}$ are centered
on $x$ and $x'$, respectively.  The goal here is to study the half-filled
Hubbard model by evaluating $Dm_0$ and $Dm_1$ in the two-site approximation.

Because the procedure for implementing the two-site
approximation has been described previously\cite{mm,mm1}, we will only outline the
essence of 
the method: 1) enumerate the quantum mechanical states for two neighbouring
sites, 2) use
the resolvent method\cite{kuramoto} to express how the surrounding environment
interacts with the two-site system, 3)
express the $\delta j's$ in terms of the level operators for the two-site
system, and 4) use the non-crossing approximation to expand $Dm(\vec k,\omega)$
in terms of the two-site states.  For two sites,
there are $4^2=16$ electronic states.  Let $B(i)$, $F_\sigma(i)$, and $D(i)$, represent
single site level operators acting on
empty (a boson state), singly (with spin $\sigma$) and doubly
occupied sites respectively. In terms of the level operators,
the original Hubbard operators are
$\xi_\sigma=B^\dagger F_\sigma$ and $\eta_\sigma=\sigma 
F^\dagger_{-\sigma}D$.  The level operators for
two-site states, $\Phi_n$,
are formed by all possible symmetric and antisymmetric combinations of $B$,
$F_\sigma$ and $D$ on two neighbouring sites. Two particular states
of interest are the singlet
\beq\label{sing} 
FF_A=\frac{F_\uparrow F^\prime_\downarrow-F_\downarrow F^\prime_\uparrow}
{\sqrt{2}}
\eeq
and triplet state formed from three states,
\beq\label{trip}
FF_S=\frac{F_\uparrow F^\prime_\downarrow+F_\downarrow F^\prime_\uparrow}
{\sqrt{2}},
\eeq
as well as $F_{\uparrow}F^\prime_{\uparrow}$ and $F_{\downarrow}F^\prime_{\downarrow}$.
The two-site system described by all
 sixteen states of this form is not isolated.  The rest of the system
acts as a reservoir.  The coupling is realized 
through the electron propagator for the rest of the 
system, which can be obtained from the appropriate trace
of the single-particle Green function defined in Eq. (\ref{fgf}).  
Consequently, the energy levels of the two-site states are broadened.
We define then the resolvent of a two-site state, 
$R_{nm}(t)=\theta(t)\langle\langle\Phi_n(t)\Phi^\dagger_m(0)\rangle\rangle$, as a trace over 
the degrees of 
freedom of the reservoir and the two-site subsystem.  The Fourier transform of the resolvent
\beq
R_{nm}=\int d\omega'\frac{\sigma_{nm}(\omega')}{\omega-\omega'+i\delta}.
\eeq
can be written in terms of the spectral function for the two-site
system, which is defined as $\sigma_{nm}=-{\rm Im}R_{nm}/\pi$.
The $\delta j$'s are then expanded
\beq
\delta j=\sum_{nm} a_{nm} \Phi^\dagger_n\Phi_m
\eeq
in terms of the sixteen level operators that describe the two-site physics.
The products of $\delta j's$ that comprise the dynamical
corrections are then expressed in terms of the two-site
resolvents using the non-crossing approximation (NCA). Consequently,
the total self-energy contains all the possible convolutions
of the two-site resolvents.  As the dynamics included are local,
the NCA is expected to be accurate\cite{kuramoto}.
The resolvents 
associated with local triplet and singlet states
are sharply peaked at well-defined energies\cite{mm}.  The energy difference
is small, however, and given roughly by $t^2/U$.  Consequently,
singlet-triplet mixing cannot be ignored.  We can introduce this
spin-fluctuation effect as a correction to the self-energies of 
the resolvents.  The effective energy of the exchange interaction is
given by the difference,
\beq\label{JEFF}
J\equiv \int_{-\infty}^{\infty} \omega\left(\sigma_{FF_S}-\sigma_{FF_A}
\right)d\omega,
\eeq
 of the first moments of the spectral functions for the singlet and triplet
two-site states.
Antiferromagnetism corresponds to $J>0$.
To facilitate a self-consistent
evaluation of the exchange energy, we computed
the equations of motion for the two-site level operators
in the presence of the spin-fluctuation term, $J \vec n\cdot\vec n^\alpha/2$.
Here ${\vec n}=(c_{i\uparrow},c_{i\downarrow})\sigma_i
(c_{i\uparrow},c_{i\downarrow})^\dagger$, with $\sigma_i$ the Pauli matrices
and $\alpha$
represents an average over the nearest neighbours of the second site in the cluster
exclusive of the first site.
A final quantity that is needed is the occupancy in the two-site states
\beq\label{nocc}
n_{\Phi_n}=Z_{\Phi_n}/Z,
\eeq
where $Z_{\Phi_n}=\int d\omega \bar{\sigma}_{nn}(\omega)$,
 where $Z$ represents
a sum over all $Z_{nm}$'s for the sixteen states and
$\bar{\sigma}_{nm}=exp(-\beta\omega)\sigma_{nm}(\omega)$. 

\section{Results}

To obtain a self-consistent solution for the self-energy, we start with 
an initial guess for the electron spectral function which will
be used to describe the properties of the environment in
which the two-site system
is placed.  The two-site resolvents are then determined iteratively.
Once the resolvents are determined, they are fed into the
dynamical corrections and the Green function is determined. A new spectral function
is then computed thereby closing the self-consistent set of equations.
This procedure is repeated until convergence is reached.

\subsection{Hubbard gap}

\begin{figure}
\begin{center}
\epsfig{file=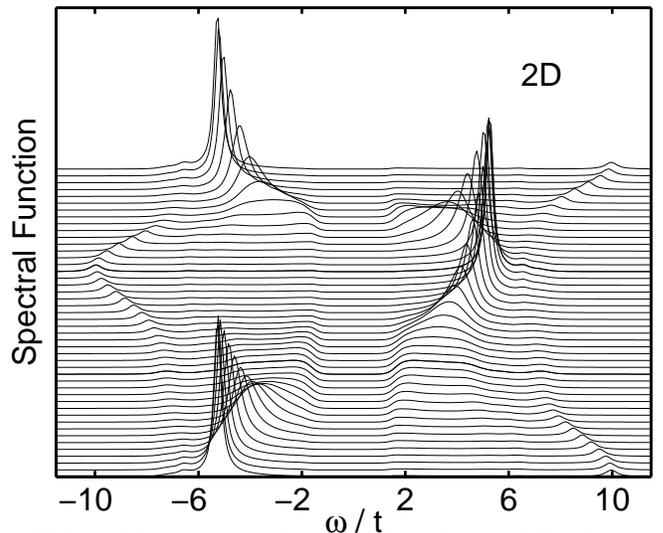, height=7cm}
\caption{Momentum and energy dependence of 
the electron spectral function for $U=8t$ and $T=0.15t$ for
different values of $\vec k$.  From top to bottom,
the momenta correspond to $(k_x,k_y) = (0,0)\rightarrow
(\pi,\pi)\rightarrow (\pi,0)\rightarrow (0,0)$.  Each
momentum trace is shifted by hand.}
\label{fig1}
\end{center}
\end{figure}

\begin{figure}
\begin{center}
\epsfig{file=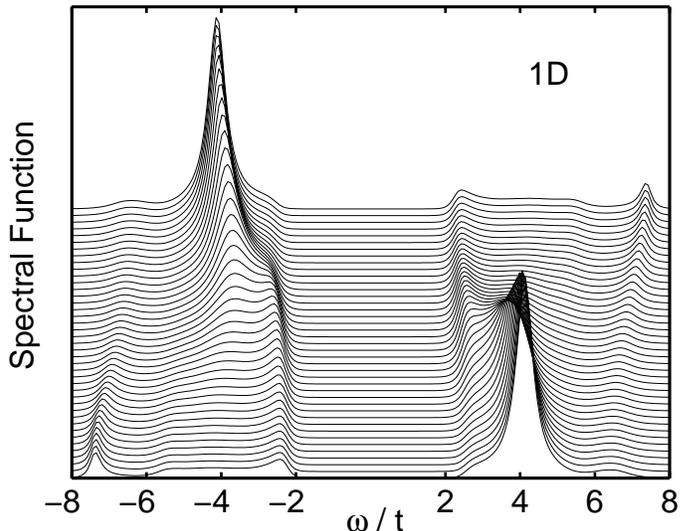, height=7cm}
\caption{Momentum and energy dependence of 
the electron spectral function for $U=8t$ and $T=0.15t$ for
different values of $\vec k$.  From top to bottom,
the momenta correspond to $k_x = 0\rightarrow\pi$.  Each
momentum trace is shifted by hand.}
\label{hu1}
\end{center}
\end{figure}

Shown in Fig. (\ref{fig1}) is the total electron spectral function
$(- {\rm Im} (G_{11}(k,\omega)+2G_{12}(k,\omega)+G_{22}(k,\omega))/\pi)$ for the half-filled
Hubbard model for $U=8t$ and $T=0.15t$.
Each trace corresponds to a momentum
starting from $(0,0)$ to $(\pi,\pi)$ to $(\pi,0)$ and then back
to $(0,0)$.   Clearly shown are the upper and lower Hubbard bands with an
 energy gap of order $U$ and the flatness of the band near the
$(\pi,0)$ point.  The broadening is due entirely to the dynamical corrections.
In the absence of such processes, the spectral function would correspond to a
series of $\delta$-functions at the lower and upper Hubbard
bands.  In the upper Hubbard band,
the dominant spectral weight lies at the $(\pi,\pi)$ point whereas in
the lower band the spectral weight is located at $(0,0)$. 
Broad spectral features obtain near the
 $(\pi,0)$ point.  The analogous spectral function
for the D=1 case is shown in Fig. (\ref{hu1}). 
As in the 2D case, the energy separation between the lower and upper Hubbard
bands for D=1 is striking. In both cases, the periodicity of the band is
$2\pi$ as is expected for a paramagnetic solution.  Nonetheless, we will 
show that local antiferromagnetic correlations
exist.

\begin{figure}
\begin{center}
\epsfig{file=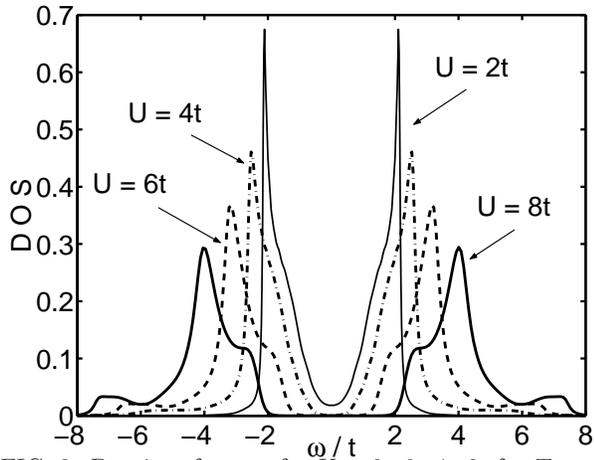, height=6cm}
\caption{Density of states for $U=8t,6t,4t,2t$ for $T=.15t$ in D=1.  
The presence of a
gap for all values of $U$ indicates an absence of a metallic state at half-filling.}
\label{dos1}
\end{center}
\end{figure}

\begin{figure}
\begin{center}
\epsfig{file=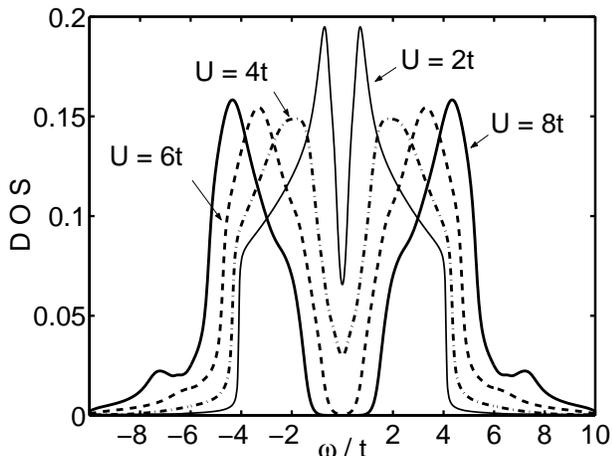, height=6cm}
\caption{Density of states for $U=8t,6t,4t,2t$ for $T=.15t$ in $D=2$.
The presence of a
gap for all values of $U$ indicates an absence of a metallic state at half-filling.}
\label{dos2}
\end{center}
\end{figure}

\begin{figure}
We see clearly that $\rho(0)$ is non-zero even in the strong-coupling regime, 
$U\gg t$,  provided that
the temperature is sufficiently large.  This signifies that the physics
at small and large
$U$ do not differ qualitatively.   Because we have probed both $U<W$ and $U>W$, we conclude
that the Hubbard gap persists for all values of $U$ both in $D=1$ and $D=2$
and is certainly fully formed at $T=0$.  Hence, the only signular point in the 
half-filled Hubbard model is $U=0$ where the gap disappears.  Consequently,
there is an absence of a metallic state at half-filling.   
This result is in agreement with that of Moukouri and Jarell\cite{dca} and is consistent with the argument
of Anderson\cite{anderson} that the Hubbard model is always in the
strong-coupling regime.
\begin{figure}
\begin{center}
\epsfig{file=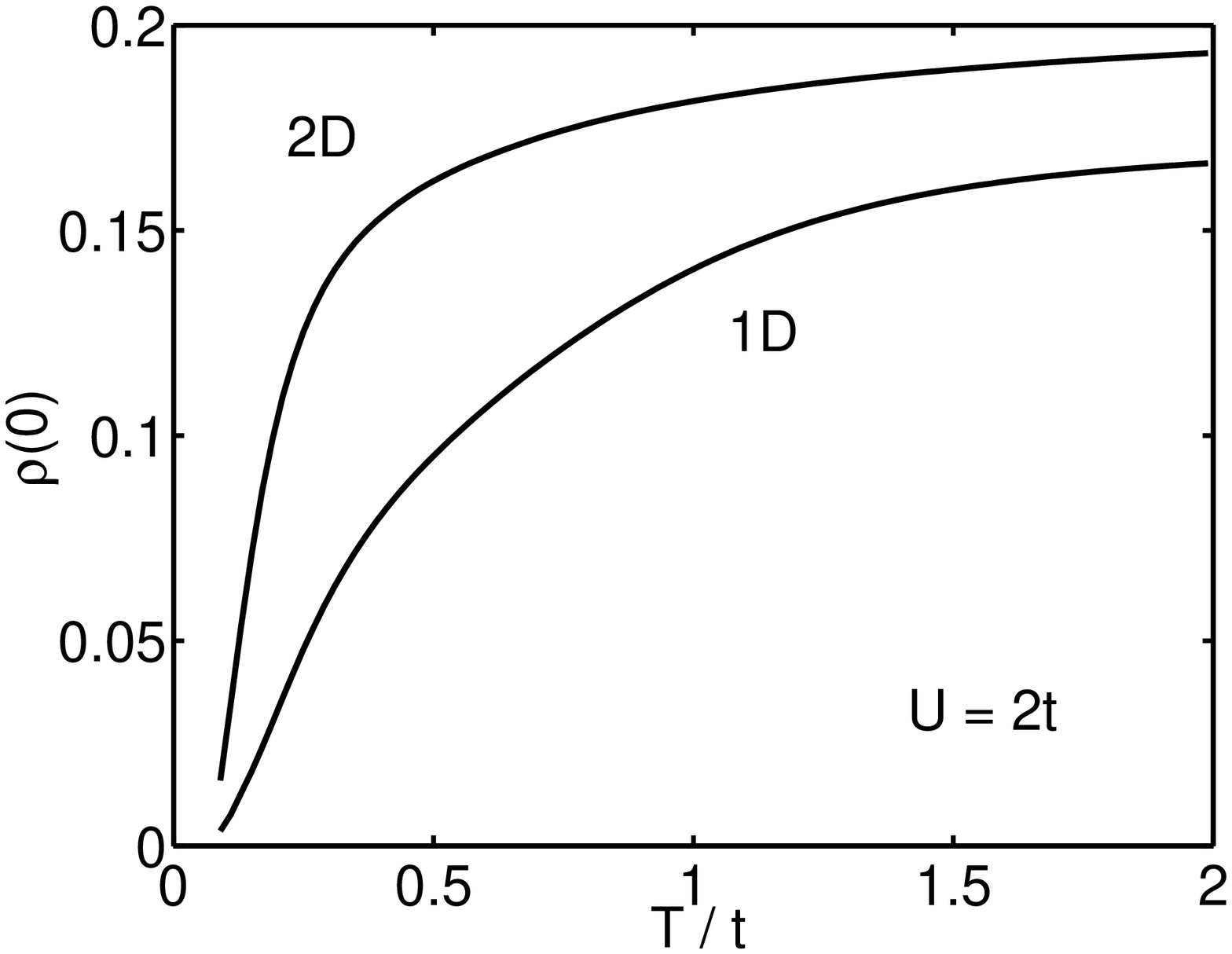, height=6cm}
\caption{Density of states at the Fermi energy for $U=2t$ as 
a function of temperature.}
\label{tdep1}
\end{center}
\end{figure}
\begin{center}
\epsfig{file=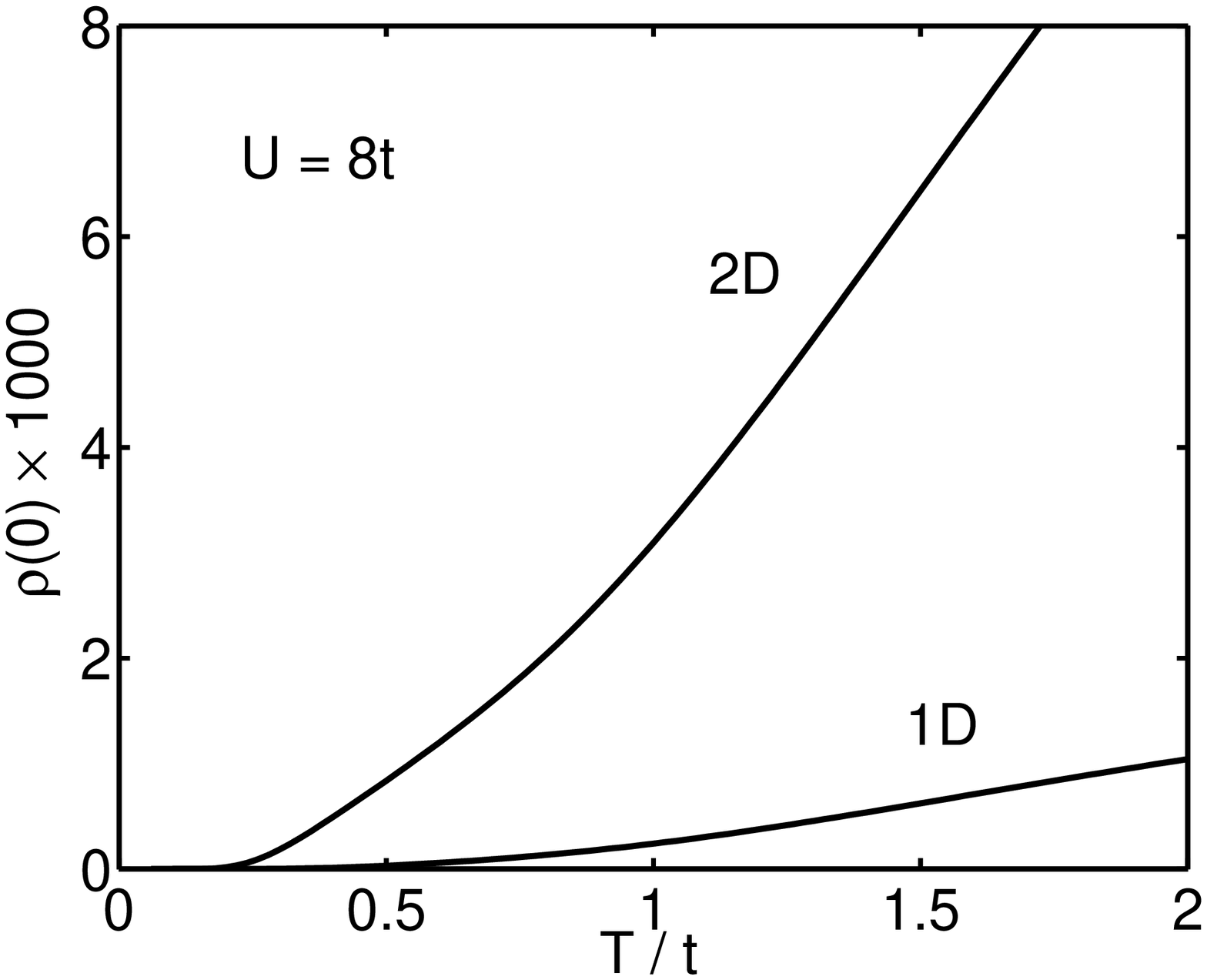, height=6cm}
\caption{Density of states at the Fermi energy for $U=8t$ as 
a function of temperature.}
\label{tdep2}
\end{center}
\end{figure}

\begin{figure}
\begin{center}
\epsfig{file=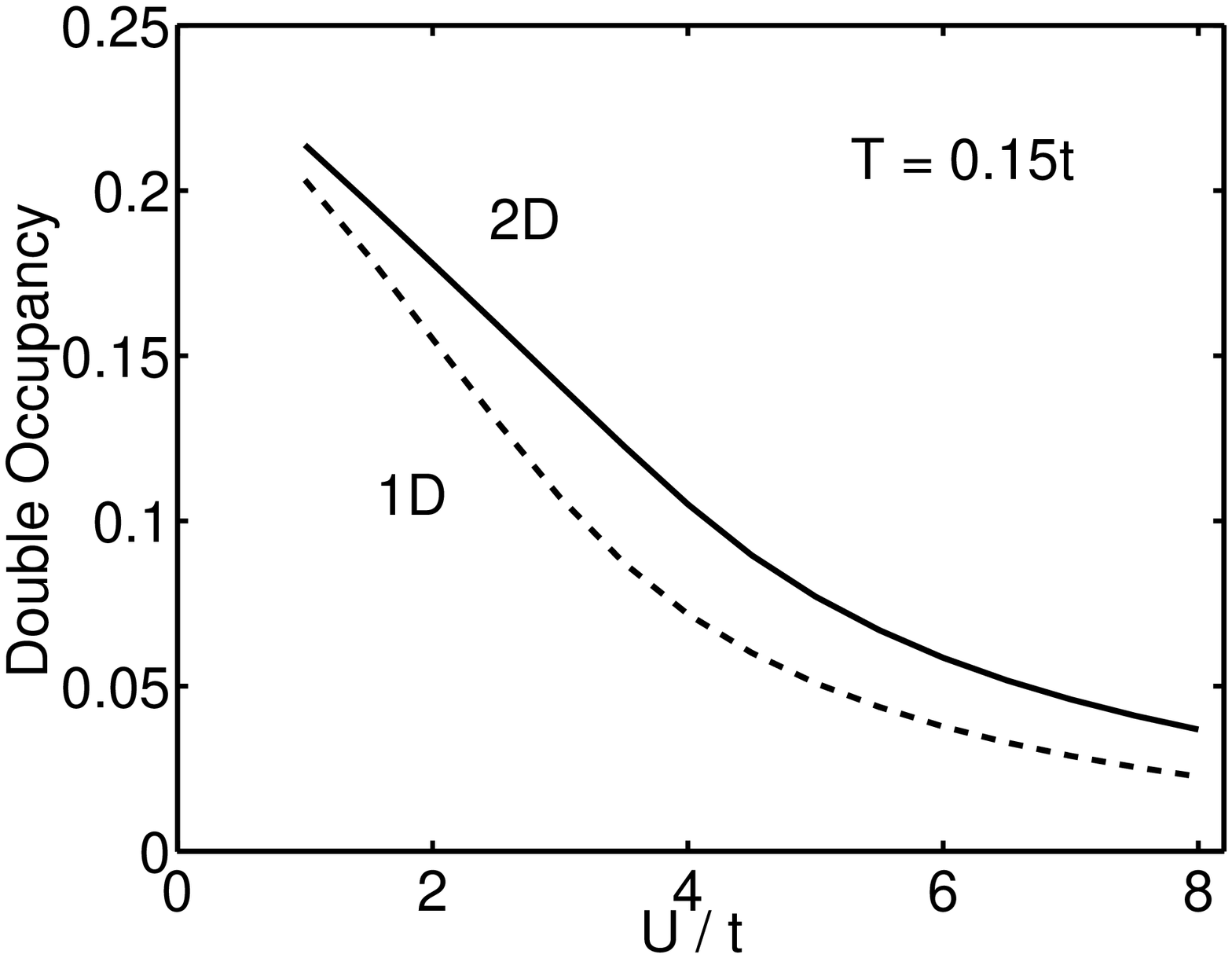, height=6cm}
\caption{Double occupancy as a function of $U$.  In both D=1 and D=2, 
the double occupancy approaches the non-interacting value of 
$1/4$ as $U\rightarrow 0$ and vanishes as $U\rightarrow\infty$.}
\label{docc}
\end{center}
\end{figure}

Integration of the spectral functions with respect to momentum yields the density of states (DOS) as a function
of energy shown in Figs. (\ref{dos1}) and (\ref{dos2}).
\noindent It is evident that
the Hubbard gap is fully formed for $U\approx W$, where $W$ is
the bandwidth ($W=4t$ for $D=1$ and $W=8t$ for $D=2$). In the D=1 case,
the gap is wider than in $D=2$. For all values of $U$, we see
a clear suppression in
the density of states near zero energy.   However, for small
values of $U$, the density of states at the Fermi
energy does not vanish at the temperatures we consider here.  
It is crucial then that we investigate the temperature dependence
of the density of
states at zero energy for small $U$.  Shown in Fig. (\ref{tdep1})
is the density of states at the Fermi energy, $\rho(0)$ as a function
of temperature  
for both $1D$ and $2D$ at $U=2t$. 
For both $D=1$ and $D=2$, $\rho(0)$ drops to zero as the temperature
decreases.  The enhancement seen in the density of states at zero energy
in the $D=2$ case over the $D=1$ problem is tied to the shape of 
the non-interacting density of states.  In D=1, as $U$ decreases, the density
of states becomes sharp (and actually diverges) at the band edges rather than 
approaching a constant
value as for $D=2$.  In fact, in $D=2$, the band exhibits a singularity at 
the gap edge.  Consequently, it is easier to fill in the gap in D=2 than
in D=1 as is seen in Fig. (\ref{tdep1}).  This trend also persists 
for any value of $U$ as demonstrated in Fig. (\ref{tdep2}) for $U=8t$.
Additional confirmation of the continuity between the weak and strong coupling
regimes at half-filling is seen from the double occupancy shown in 
Fig. (\ref{docc}). In terms of the two-site resolvents, the
expression for the double occupancy is given by
\beq
D=\frac12(n_{FDS}+n_{FDA}+n_{DBS}+n_{DBA})+n_{DD}
\eeq
where $n_\phi$ is given by Eq. (\ref{nocc}). In both cases, the double 
occupancy smoothly approaches the non-interacting value of $1/4$ as $U$
decreases.  The absence of any kink in the double
occupancy indicates an absence of a phase transition between the regimes,
$U<W$ and $U>W$.  Hence, a continuity appears to exist between the small
and large $U$ regimes in the half-filled Hubbard models.  The existence
of a non-perturbative gap indicates that half-filled $1D$ and $2D$ Hubbard 
models flow to strong coupling.

\subsection{Local Antiferromagnetism}

\begin{figure}
\begin{center}
\epsfig{file=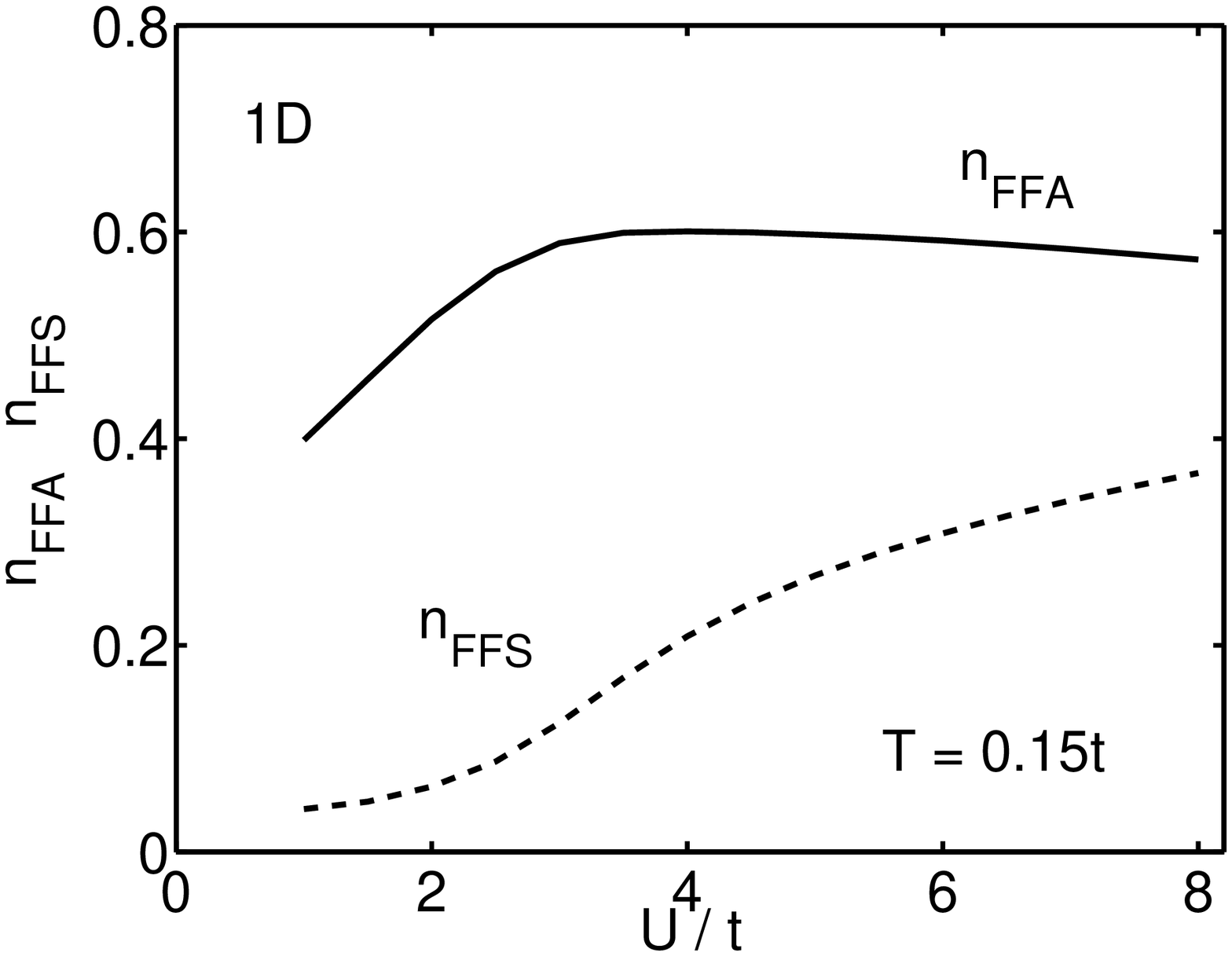, height=6cm}
\caption{Singlet ($n_{FFA}$) and triplet ($n_{FFS}$) state occupancies
as a function of $U/t$ for D=1.  For any non-zero value of $U$,
$n_{FFA}>n_{FFS}$ indicating a tendency toward antiferromagnetic
order.}
\label{occ1d}
\end{center}
\end{figure}

\begin{figure}
\begin{center}
\epsfig{file=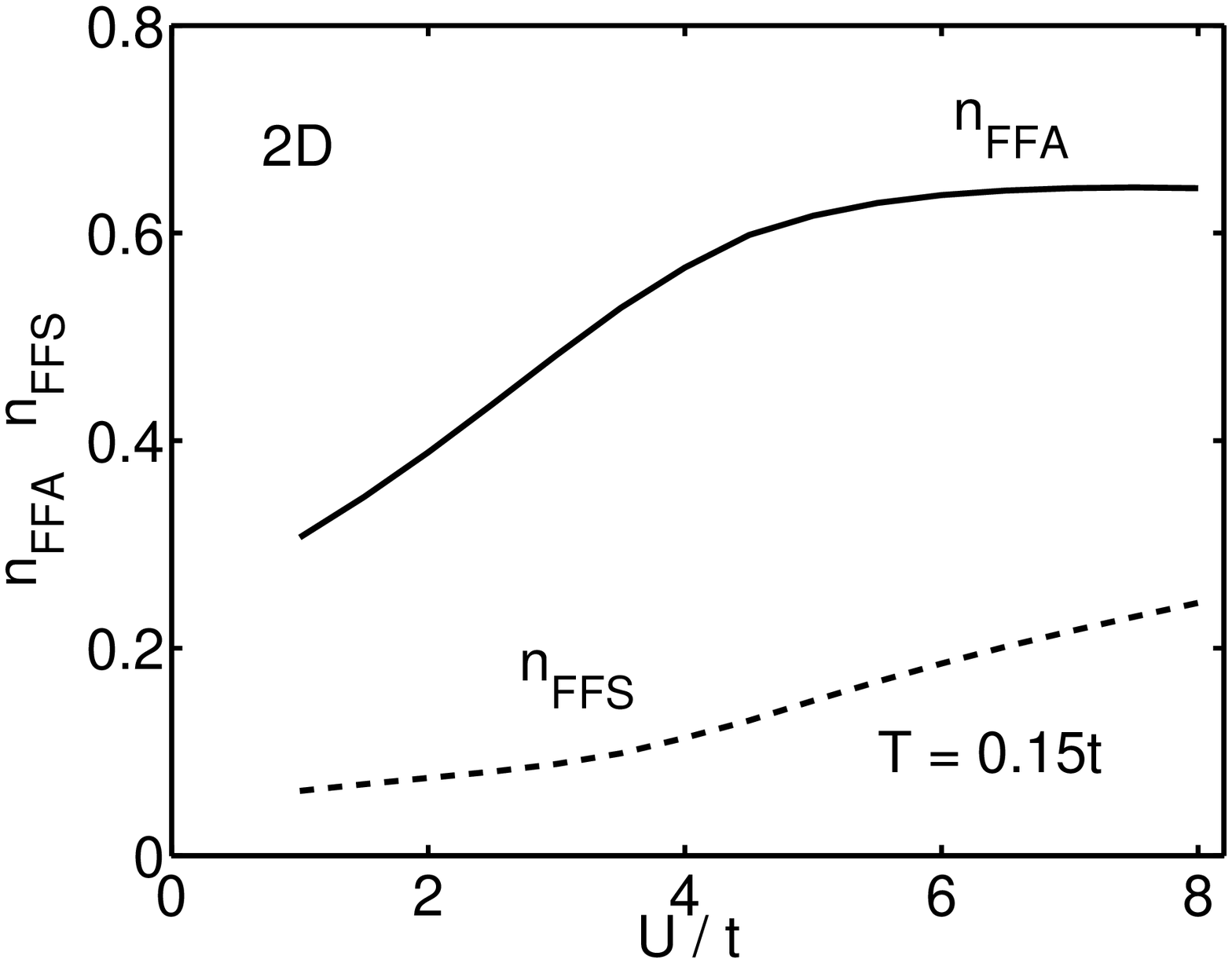, height=6cm}
\caption{Singlet ($n_{FFA}$) and triplet ($n_{FFS}$) state occupancies
as a function of $U/t$ for D=2. For any non-zero value of $U$,
$n_{FFA}>n_{FFS}$ indicating a tendency toward antiferromagnetic
order.}
\label{occ2d}
\end{center}
\end{figure}

Because our method is capable of uncloaking the underlying spin dynamics, we 
investigated the behaviour of the local triplet, $n_{FFS}$, and 
singlet, $n_{FFA}$, occupancies for the states defined
in Eqs. (\ref{trip}) and (\ref{sing}), respectively.
These quantities were computed directly from the resolvents that
enter  Eq. (\ref{nocc}).  Should
the ground state be an antiferromagnet, the singlet occupancy, $n_{FFA}$,
should exceed the triplet occupancy, $n_{FFS}$ at sufficiently
low temperatures.  For both 1D and 2D the plots of 
the singlet and triplet occupancies are shown in Figs. (\ref{occ1d}) and
(\ref{occ2d}), respectively.

\begin{figure}
\begin{center}
\epsfig{file=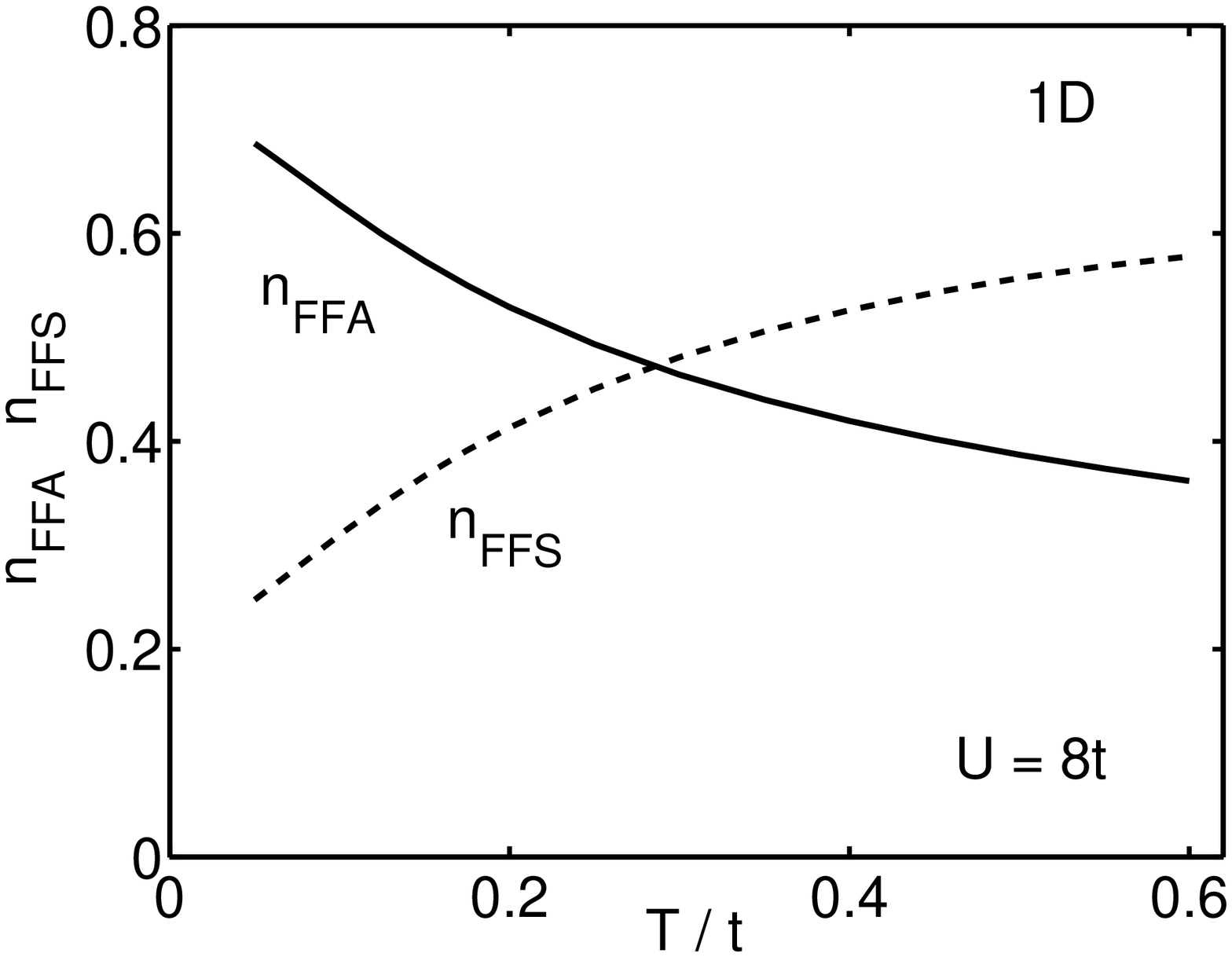, height=6cm}
\caption{Singlet ($n_{FFA}$) and triplet ($n_{FFS}$) state occupancies
as a function of temperature for D=1.  The fact that the singlet
occupancy as $T\rightarrow 0$ becomes of order unity
is consistent with antiferromagnetic order.}
\label{noc1d}
\end{center}
\end{figure}

\begin{figure}
\begin{center}
\epsfig{file=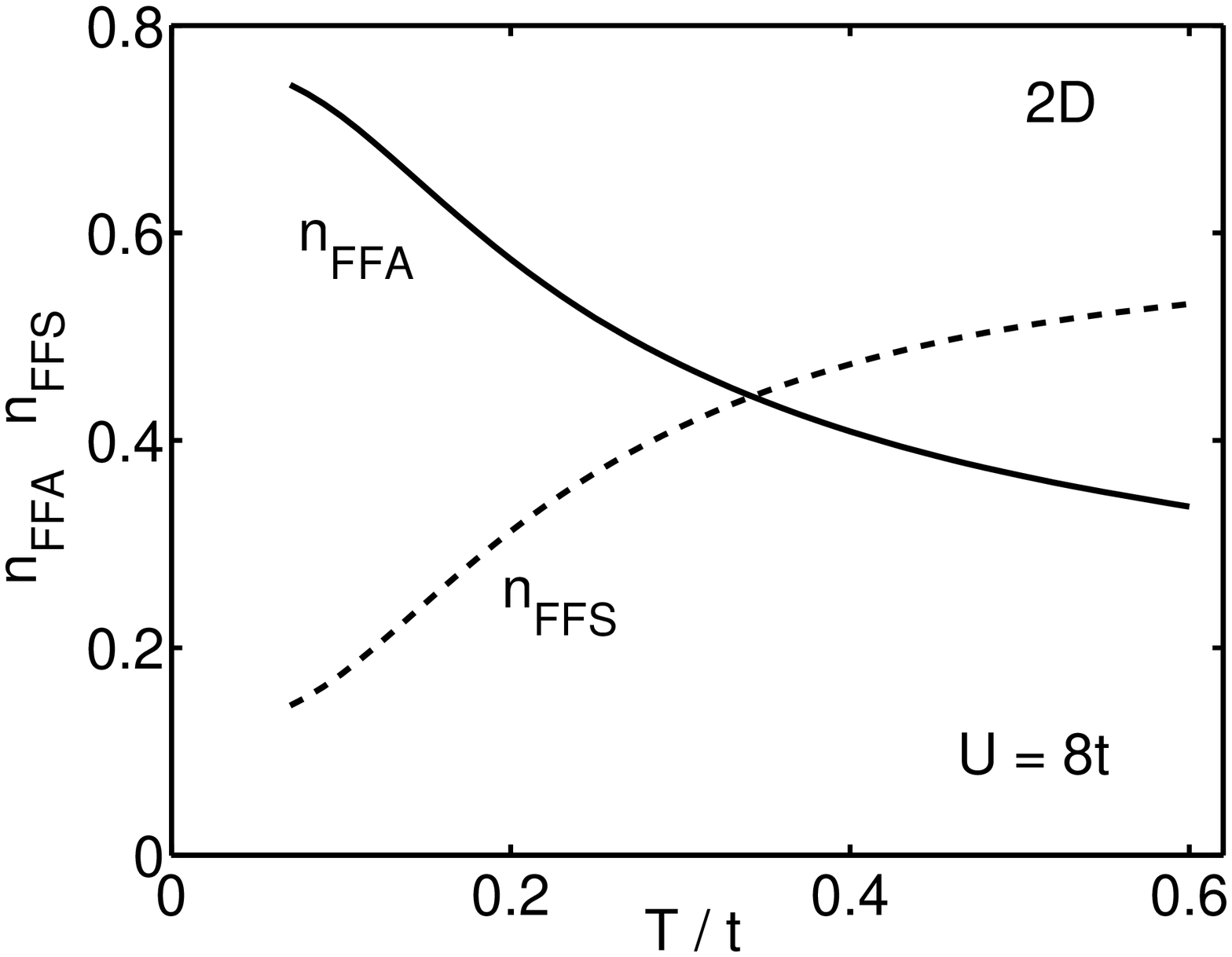, height=6cm}
\caption{Temperature dependence of the singlet ($n_{FFA}$) and triplet
($n_{FFS}$)
occupancies for D=2.  The fact that the singlet
occupancy as $T\rightarrow 0$ becomes of order unity
is consistent with antiferromagnetic order.}
\label{noc2d}
\end{center}
\end{figure}

As is evident, $n_{FFA}>n_{FFS}$ as is consistent with an ordering tendency
toward antiferromagnetism.  Notice in the large $U$ limit in $2D$, singlet state formation
is enhanced over the 1D value.  However, to ascertain if these local probes
are true signatures of ground state properties, we computed the temperature
dependence of $n_{FFS}$ and $n_{FFA}$ for both 1D and 2D at $U=8t$.
From Figs. (\ref{noc1d}) and (\ref{noc2d}), we find thatat high temperatures
triplet excitations dominate.  However, this trend is reversed below some
temperature and the singlet occupancy becomes of order unity.
This is significant and implies that the ground state is in fact
an antiferromagnet.  The tendency toward antiferromagnetism appears
to be slightly enhanced in 2D relative to the 1D problem.
Our results then are consistent with antiferromagnetic order at $T=0$ in 
both 1D and 2D.

\begin{figure}
\begin{center}
\epsfig{file=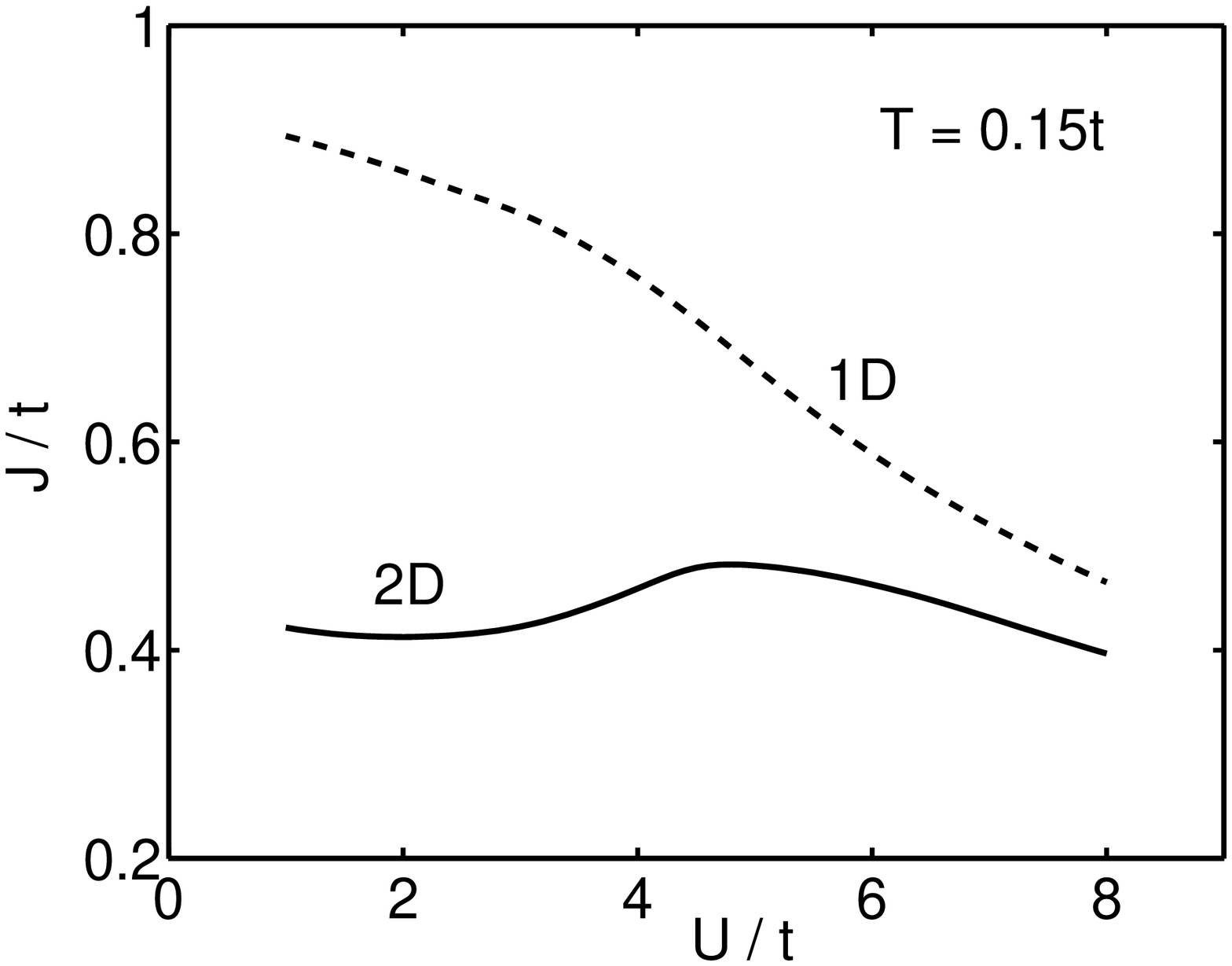, height=6cm}
\caption{Effective exchange interaction $J$ as a function
of $U/t$ computed using Eq. (\protect\ref{JEFF}).}
\label{J}
\end{center}
\end{figure}

The energy splitting between the singlet and triplet states is due to an
effective exchange interaction. Using Eq. (\ref{JEFF}), we computed
the effective exchange interaction shown in Fig. (\ref{J}) for both $1D$ and
$2D$.  Note first that $J$ is always positive as a consequence of the fact
that the singlet state is lower in energy than the triplet. This is a further
indication of the antiferromagnetic order in the ground state.
As expected, $J$ is well approximated
by $t^2/U$ in the strong-coupling regime.  However,
as $U$ decreases, deviations from this behaviour are observed. 

\subsection{Heat Capacity}

\begin{figure}
\begin{center}
\epsfig{file=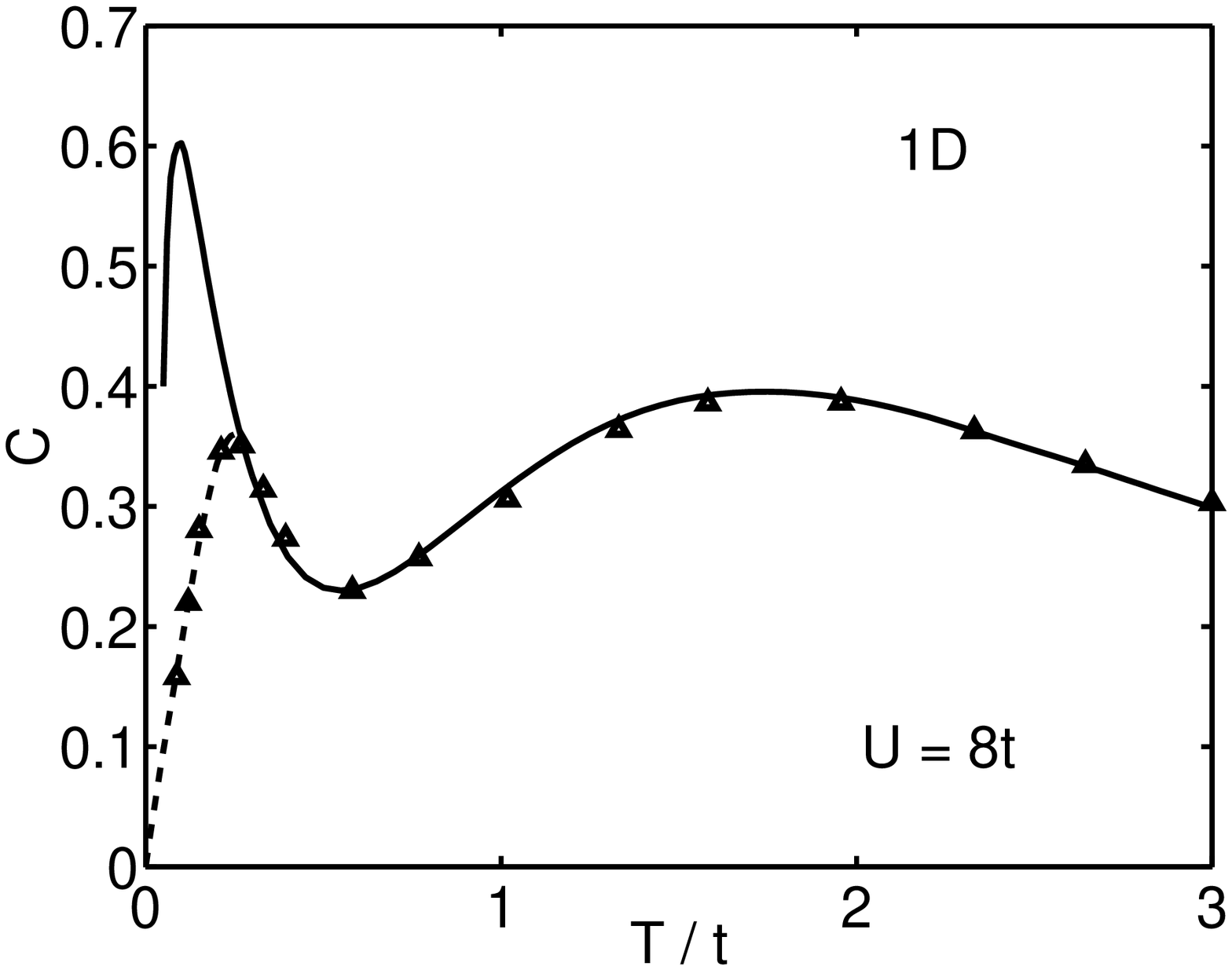, height=6cm}
\caption{Heat capacity for the 1D half-filled Hubbard model
as a function of temperature for $U=8t$.  The filled triangles are the 
results from the exact treatment via Bethe
 ansatz (see Ref. (\protect\cite{bethe})).}
\label{heat1du8}
\end{center}
\end{figure}

\begin{figure}
\begin{center}
\epsfig{file=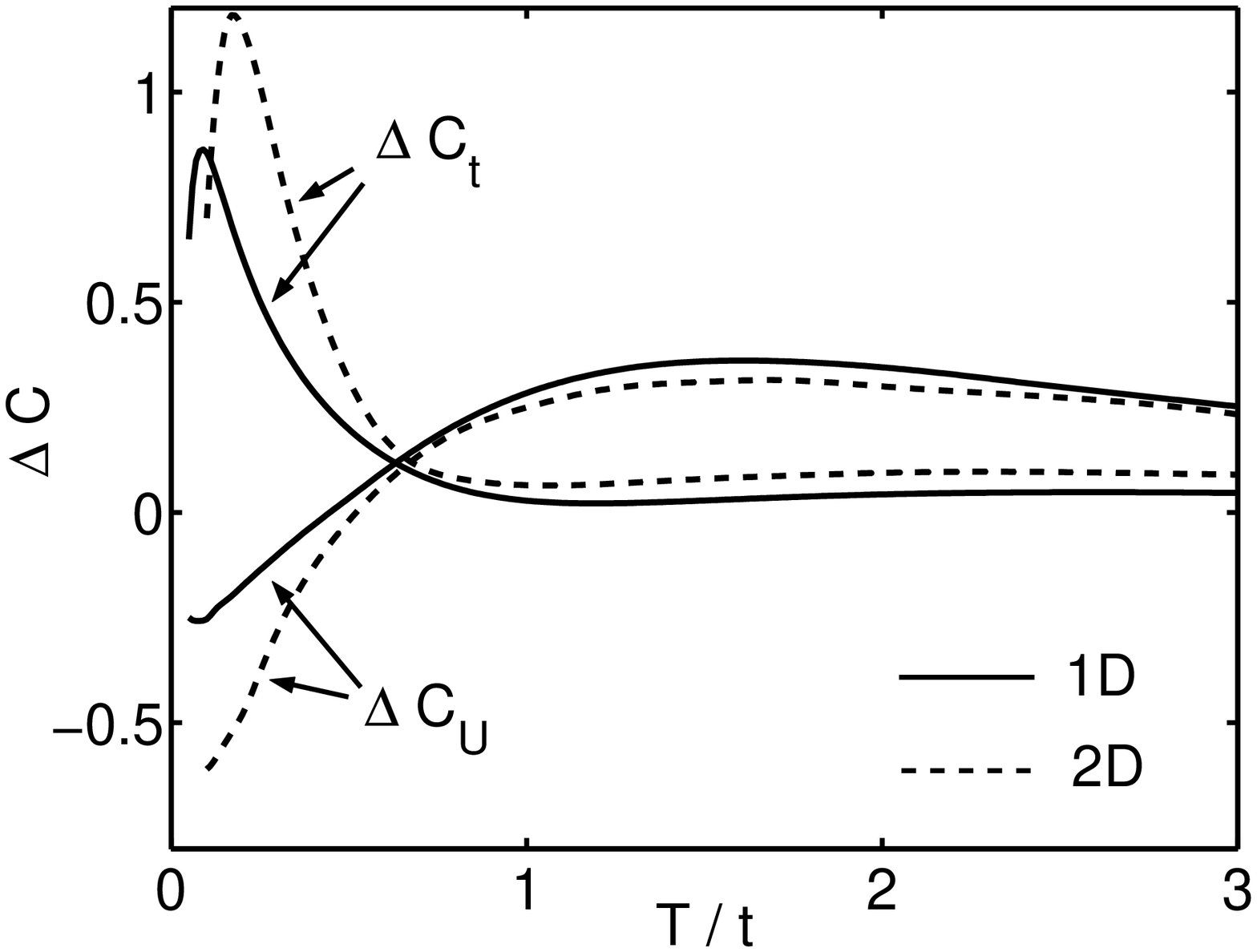, height=6cm}
\caption{Kinetic ($\Delta C_t$) and potential ($\Delta C_U$)
 energy contributions
to the heat capacity of the 1D and 2D half-filled Hubbard models
for $U=8t$.}
\label{cku}
\end{center}
\end{figure}

\begin{figure}
\begin{center}
\epsfig{file=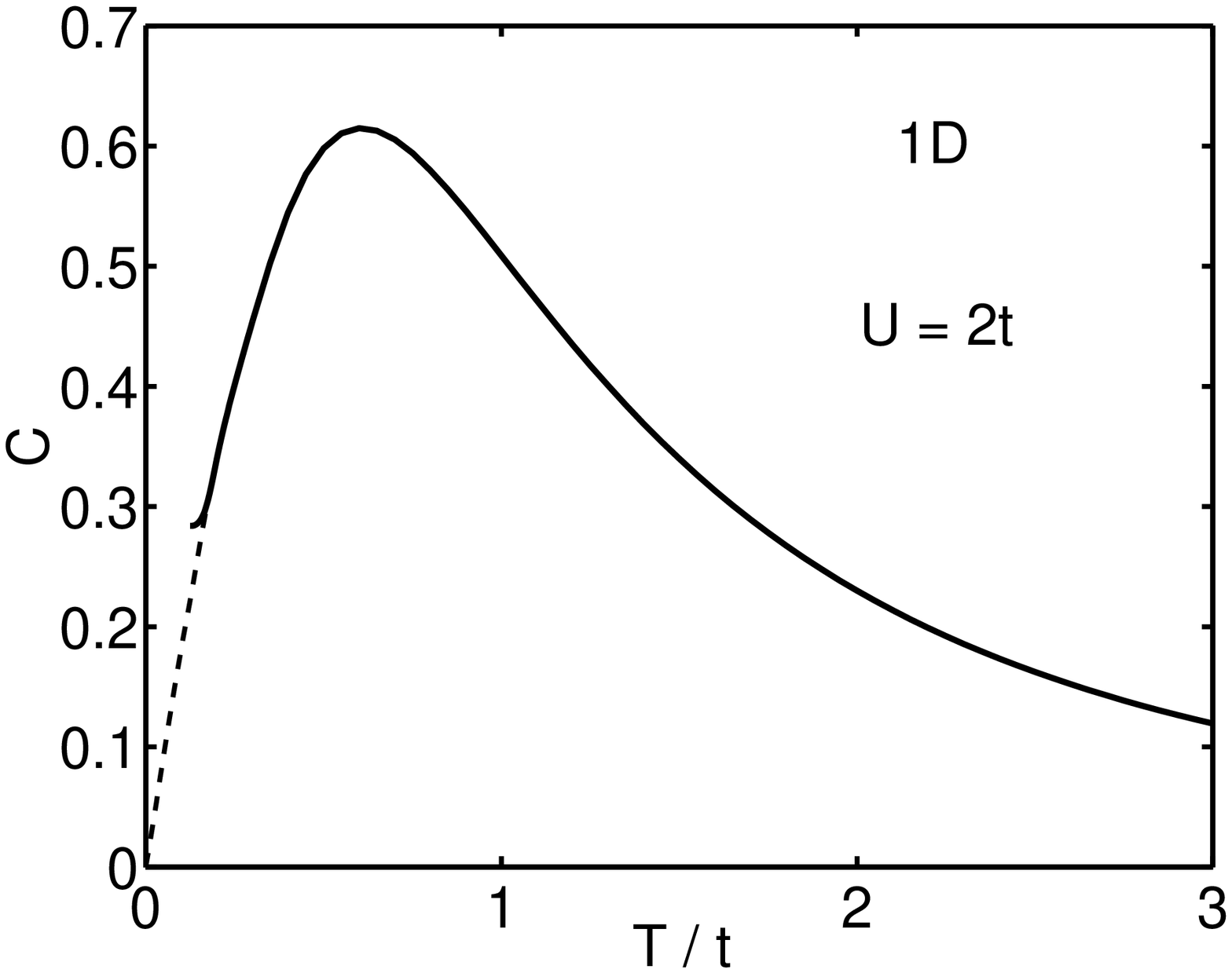, height=6cm}
\caption{Heat capacity for the 1D half-filled Hubbard model
as a function of temperature for $U=2t$.  The deviation
from the dashed line at low temperature stems from the emergence of sharp spectral
features below $T\approx 0.1t$ which prohibit an accurate numerical
determination of the integral that enter the heat capacity.}
\label{weakheat}
\end{center}
\end{figure}

The natural question that arises with any approximate treatment
of strong correlation physics is: How seriously should the results
be taken?  Our study of the 1D problem is in part motivated
by the fact that exact results are available
from the Bethe ansatz.  While Bethe ansatz is not amenable to 
yielding the Green functions from which the density of states can be
calculated, ground state energies and thermodynamic quantities
are readily available by this technique.  Rather than compare with the total
energy, we compute the temperature derivative or the heat capacity.  Computation
of the average energy is straightforward because we have already obtained the 
average double occupancy.  Shown in Fig. (\ref{heat1du8}) is a comparison
between the heat capacity computed within the present method (solid line) and the prediction
from the Bethe ansatz\cite{bethe} (triangles) for $U=8t$. 
This figure demonstrates
that at high to moderately low temperatures,
the present method is quantitatively accurate, yielding results which differ by no more
than $1\%$ from those of the Bethe ansatz. Such agreement is significant because
in 1D, correlation effects are particularly amplified.  The two-peak
structure of the heat capacity is tied to a competition between the 
contribution from the potential energy (high $T$) and 
the kinetic energy (low $T$) as illustrated in Fig. (\ref{cku}). 
Both display
maxima but in distinctly different energy regimes.  
Near perfect agreement with the Bethe ansatz solution is obtained
at high temperatures where the potential energy dominates.  This is
to be expected as the Hubbard operators provide an accurate treatment of
the potential energy but only an approximate description of the
kinetic energy.  At sufficiently 
low temperatures,
where the kinetic energy dominates, sharp spectral features appear 
and the numerical accuracy of the method wanes. 
Another source of error
could be the two-site approximation itself.  At low temperatures,
an accurate description of the low-energy physics becomes essential.
It might be that the two-site approximation inherently over-estimates the 
magnitude of the kinetic energy.  To see if this breakdown persists
for small $U$, we computed the heat capacity
for $U=2t$.  The two-peak structure
that occurs in the large U regime is absent for $U\ll W$ as illustrated
in Fig. (\ref{weakheat}).  The disappearance of the two peaks
is dictated by the non-interacting limit which possesses a single peak
at $T\approx 0.5t$.  Our results are in quantitative agreement
with the numerical simulations of Shiba and Pincus\cite{sp} 
down to $T\approx .1t$.
Below this temperature, lack of numerical precision prohibited any
accurate determination of the heat capacity.  It appears then that the source
of the breakdown at low temperatures stems more from the lack of numerical
 accuracy
than from the local description of the physics.  However, more studies on this are necessary.

\begin{figure}
\begin{center}
\epsfig{file=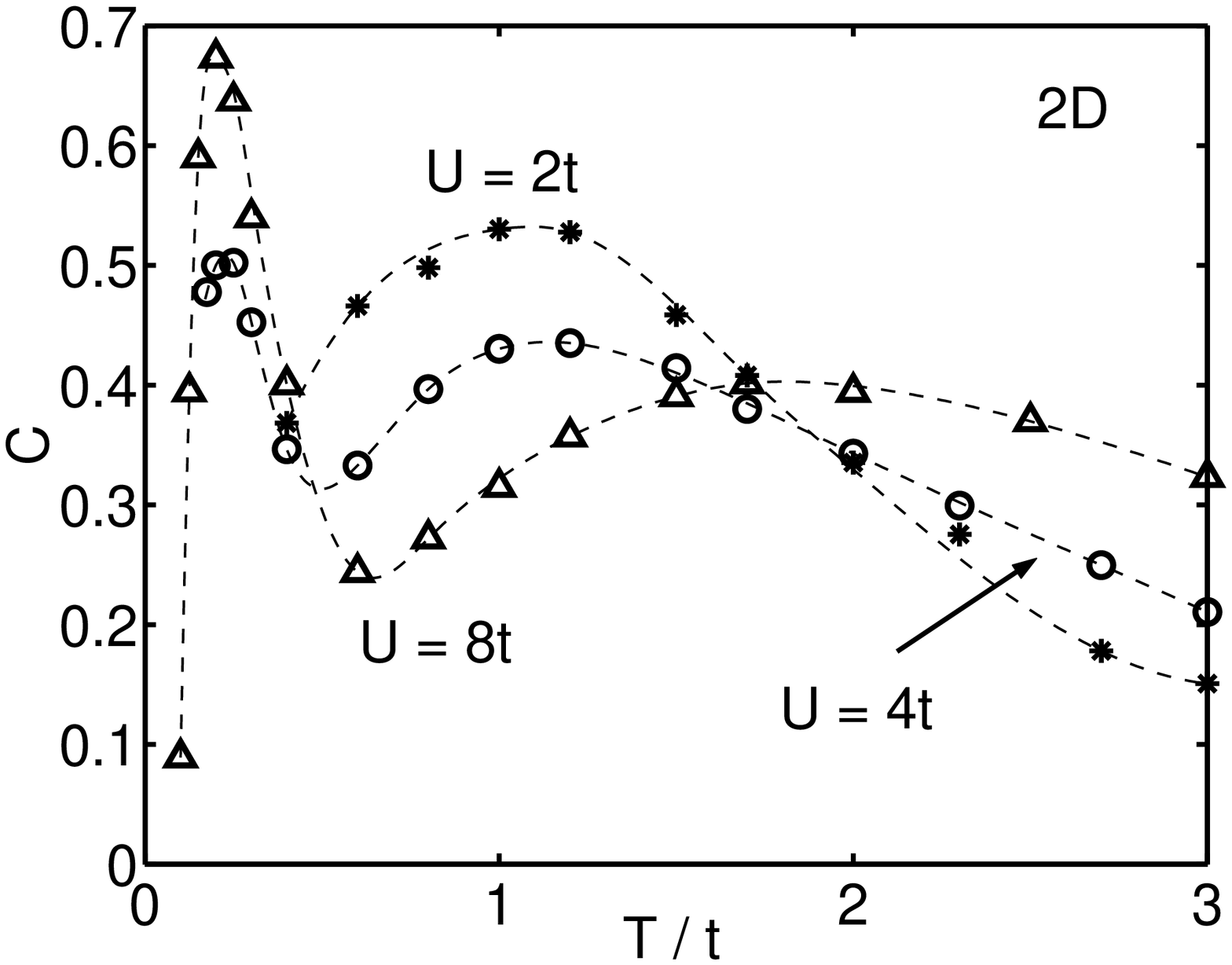, height=6cm}
\caption{Heat capacity for the 2D half-filled Hubbard model
as a function of temperature for $U=2t, 4t, 8t$.  The crossing
points at $T\approx 1.7t\pm 0.1t$ and at $T=.4t\pm 0.1t$ is in
agreement with the general arguments of Ref. (\protect\cite{voll}).
The low-T peak in the heat capacity arises from spin fluctuations and the 
high temperature physics is tied to charge fluctuations.}
\label{heat2d}
\end{center}
\end{figure}

In 2D, the heat capacity (see Fig. (\ref{heat2d})) has the 
familiar 2-peak structure of the 1D problem.  Here again,
this structure arises from a competition between the kinetic and potential
energies as the dashed lines in Fig. (\ref{cku}) reveal. Similar results
have also been obtained by Scalettar and colleagues~\cite{scall} from
quantum Monte Carlo simulations on finite samples. The final
feature on which we focus is the crossing of $C(T,U)$ versus $T$ for various
values of $U$.
In $1D$ comparison of Fig. (\ref{heat1du8}) with (\ref{weakheat}) reveals
that the heat capacities cross both at high temperature and at low temperature.
Fig. (\ref{heat2d}) reveals that
that this trend persists in 2D as well. 
The high temperature crossing point occurs at roughly $T\approx 1.7t\pm 0.1t$ whereas
the low-temperature crossing point is $T\approx .4t\pm 0.1$. The errors
are due largely to the uncertainty in the data at small $U$. In the quantum
Monte Carlo studies of Duffy and Moreo\cite{dm} on a $6\times 6$ square lattice, 
similar values for the low and high
temperature crossing points were found as well. 
In the $D\rightarrow\infty$ limit\cite{georg,voll},
two crossing points are observed as well though at substantially
smaller temperatures than in the 2D case.  A unique crossing point for
$C(T,U)$ as a function of $T$ for different values of $U$ implies that at a
particular temperature, the heat capacity is independent of $U$.  This behaviour
is observed in a wide variety of strongly-correlated
 experimental systems, such as
$^3He$\cite{he}, $CeCu_{6-x}Al_x$\cite{ce},  $Nd_{2-x}Ce_xCuO_4$\cite{nd},
and $UBe_{3}$\cite{ube}.  Vollhardt\cite{voll} has shown
that independence of $C(T,U)$ on $U$ at a particular temperature 
is fundamentally rooted in strong correlation physics.
 The condition for a unique crossing point for $C(T,U)$ versus $T$
for various values of $U$ can be recast\cite{voll} as
\beq\label{int}
0=\int_0^\infty\frac{dT}{T}\frac{\partial C(T,U)}{\partial U}
\eeq
in the limit that $T\rightarrow\infty$.   At high temperatures, 
$C(T,U)\propto U/T$.  
Hence, $\partial C/\partial U>0$ as $T\rightarrow \infty$.  However,
for the sum rule given by Eq. (\ref{int}) to hold, $\partial C/\partial U$ must
change sign as the temperature is lowered.  Such sign
changes will be mediated by terms proportional
to higher powers of $U$ that enter with opposing signs.
Hence, the sign change of $\partial C/\partial U$ is a true correlation
effect arising from terms at least proportional to $U^2$ and higher
in the internal energy, $E(T,U)$.  As there is no phase transition 
as a function
of temperature, the curves for $C(T,U)$ must cross to satisfy the vanishing
of the integral in Eq. (\ref{int}).   At low $T$, the width of the crossing
point is determined by low-lying excitations generated by the kinetic energy.
The natural scale for such excitations is $4t^2/U$, in rough agreement
with the low temperature crossing point in Fig. (\ref{heat2d}).  At high $T$,
charge excitations dominate the contribution to the heat capacity.
At large $U$, the gap should scale as $U-W$.
  
\section{Closing Remarks}

We have applied a local method\cite{mm,mm1}
to the determination of the dynamical corrections to the self-energy
in the single-particle Green function for the
1D and 2D Hubbard models.  Although this method focuses on local 2-site
correlations,
it captures such established features as the onset of antiferromagnetism,
absence of a Mott-Hubbard transition for non-zero $U$, and the universal
crossing of the heat capacity as a function of $T$ for various values of 
$U$.  As these features are the signatures of strong-correlation at half-filling,
it appears that local dynamics offer an adequate description of these phenomena.
Extending this method to 3 sites is prohibitive as this will entail an expansion
in $4^3=64$ three-site eigenstates.  This calculation is
impossible as the complexity of the two-site problem is already daunting.
What does seem promising, however, is a possible field theory description
of the local dynamics that seem to be essential to an accurate description
of strong correlation physics.  Work along these lines as well as extending
the present method to the doped case is underway.

\acknowledgements
We could not have completed this work without the generous help of F.
Mancini and 
H. Matsumoto who sent us previous more detailed versions of 
their work. A special thanks goes also to
H. Matsumoto who also sent us his computer program after
we had struggled several months to reproduce the
 results of Ref. (\protect\cite{mm}).  
As a result, we were able to determine
the source of the difference between our results\cite{mm1}.
We also thank the NSF grant No. DMR98-96134 for partially funding this work.

\end{document}